\documentstyle[aps,twocolumn,prl,epsf]{revtex}
\begin{document}
\twocolumn[\hsize\textwidth\columnwidth\hsize\csname @twocolumnfalse\endcsname
%%%%%%%%%%%%%%%%%%%%%%%%%%%%%%
\title{ Low-temperature coherence in the periodic Anderson model:\
Predictions for photoemission of heavy Fermions}

\author{A.\ N.\  Tahvildar-Zadeh$^1$,  M.\ Jarrell $^1$, and J.\ K.\ Freericks $^2$\\ }
\address{
         $^1$Department of Physics,
University of Cincinnati, Cincinnati, OH 45221-0011\\
         $^2$Department of Physics,
Georgetown University, Washington, DC 20057-0995\\
 	}

\date{\today}
\maketitle
\widetext
\begin{abstract}
\noindent
%        1         2         3         4         5         6         7         8
%2345678901234567890123456789012345678901234567890123456789012345678901234567890
We present numerically exact predictions of the periodic and single-impurity 
Anderson models to address   photoemission experiments on heavy Fermion 
systems.  Unlike the single impurity model the lattice model is able to account for the enhanced intensity, dispersion, and apparent weak temperature dependence of the Kondo resonant peak  seen in recent controversial photoemission experiments. We present a consistent interpretation of these results as a crossover from the impurity regime to an effective Hubbard model regime described by Nozieres.  
\end{abstract}

]

\narrowtext

\paragraph*{Introduction}

Metallic compounds containing rare earth elements with partially filled $f$ shells, such as CeBe$_{13}$ or UPt$_{3}$, belong to the general category of heavy fermions\cite{review}. They  are  characterized by a large Pauli susceptibility and  specific heat as compared to ordinary metals, which indicate a huge effective electronic mass,  and also by anomalous transport properties such as  non-monotonic temperature dependence of the resistivity. These anomalies are usually attributed to the formation of a resonant state at the Fermi energy due to the admixture of the electronically active and highly local $f$ orbitals with the metallic band of the host. Heavy fermions are usually 
modeled by the single impurity Anderson model or the periodic Anderson model
depending on the concentration of the correlated $f$ orbitals.

Photoemission experiments  provide a direct probe of the characteristic resonant states in these materials. 
However, there are two  apparently contradictory sets of photoemission results.  The large body of spectroscopic data accumulated by the Bell Lab and Neuchatel groups for scraped polycrystalline samples consistently indicate qualitative and sometimes quantitative agreement with the predictions of the single impurity models\cite{jallen}. On the other hand some recent photoemission results obtained by the Los Alamos group for single-crystalline samples cleaved {\em in situ} have revealed several apparent inconsistencies with the predictions of single impurity models, including enhanced intensity, 
a much weaker temperature dependence and a dispersion of the Kondo peak\cite{jjaa}. 
This has caused a persistent dispute regarding the ability of 
 single impurity models to  describe  the observed resonant state in  Kondo materials\cite{debate}.
 We refer the reader to  Malterre {\em{et al.}}\cite{Malterre} for a recent review. 

The single impurity Anderson model (SIAM) is typically used as the paradigm for comparison with experiments  due to its universality, the abundance of good approximations, and  the existence of exact solutions. Although at high temperatures the  SIAM  captures the same physics as the lattice model,  it cannot account for the electronic coherence at low temperatures. The  periodic Anderson model (PAM) is believed to
correctly describe the strong correlation  of $f$ electrons as well as their coherence at low temperatures and the interaction between the screened moments.

We have recently analyzed some static properties of the PAM\cite{niki2}. We found that when the $f$ band filling  $n_f\approx 1$ and the $d$ band filling $n_d \alt 0.8$ (so that the system is metallic),  the Kondo scale for the PAM, $T_0$,  is strongly suppressed compared to $T_0^{SIAM}$,  the Kondo scale for a SIAM with the same model parameters. 
However the high temperature ($T > T_0^{SIAM}$) properties of the two models are similar, so that
$T_0^{SIAM}$ is also the relevant scale for the onset of screening in the PAM whereas $T_0$ is the scale for the onset of coherence.
We found no universal relation between $T_0^{SIAM}$ and $T_0$ (their ratio depends on the $d$ band filling).
We also demonstrated that in the screening regime ($T \alt T_0^{SIAM}$) the rate of change of the local magnetic moment with temperature is smaller in the PAM than in the SIAM. 

In this paper we show that the PAM predicts a much weaker temperature dependence for the Kondo peak than the SIAM, consistent with the experiments on  single-crystalline Kondo lattice materials.  We also show that the Kondo peak is dispersive in the PAM, giving rise to heavy quasiparticle bands near the Fermi energy. We use Nozieres' idea of the effective Hubbard model for the screening clouds\cite{Nozieres} to interpret this ``band formation'' and the slow evolution of the Kondo peak. This also gives insight into 
the emergence of two relevant energy scales ($T_0^{SIAM}$ and $T_0$) in the PAM.

\paragraph*{Method} 

The PAM Hamiltonian on a $D$-dimensional hypercubic lattice is,  
\begin{eqnarray}
H &=& \frac{-t^*}{2\sqrt{D}}\sum_{\langle ij\rangle \sigma}
\left ( d^\dagger_{i\sigma}d_{j\sigma}+{\rm H.c.}\right )\nonumber \\
&+&
\sum_{i\sigma}\left(
\epsilon_{d}d^\dagger_{i\sigma}d_{i\sigma}+
\epsilon_{f}f^\dagger_{i\sigma}f_{i\sigma}
\right)
+V\sum_{i\sigma}\left(d^\dagger_{i\sigma}
f_{i\sigma}+{\rm H.c.}\right)\nonumber \\
&+&\sum_{i} U(n_{fi\uparrow}-1/2)(n_{fi\downarrow}-1/2)\;\;\label{Ham}.
\end{eqnarray}
In Eq. (1), $d(f)^{(\dagger)}_{i\sigma}$ destroys (creates) a $d(f)$ electron with
spin $\sigma$ on site $i$. The hopping is restricted to the nearest neighbors and scaled as $t=t^*/2\sqrt{D}$.  $U$ is the screened on-site Coulomb repulsion for the localized $f$ states and $V$ is the hybridization 
between $d$  and  $f$ states. This model retains the features of the 
impurity problem, including moment formation and screening, but is further complicated by the lattice effects.

Metzner and Vollhardt \cite{mevoll} observed that the irreducible 
self-energy and vertex-functions become purely local as the coordination 
number of the lattice increases. As a consequence, the solution of an 
interacting lattice model in $D=\infty$ may be mapped onto the solution of a local 
correlated impurity coupled to a self-consistently determined 
host\cite{infdrev}.  
We employ the quantum Monte Carlo (QMC) 
algorithm of Hirsch and Fye \cite{fye} to solve the remaining
impurity problem and calculate the  imaginary time local Green's functions. 
We then use  the maximum entropy method\cite{MEMlong} to find the $f$ and $d$ density of states  and  the  self-energy\cite{symmpam}.

\paragraph*{Results}

We simulated the PAM for a wide variety of fillings and parameters in units of 
$t^*$ (considered to be  a few electron-volts, the typical band-width of conduction electrons in metals).
Here we present the results for $U=1.5$, $V=0.6$, $n_f\approx 1$ with two
different $d$ band fillings $n_d=0.4$ and $n_d=0.6$ for which the Kondo scales 
are $T_0=0.014$  and $T_0=0.054$ respectively. We use the symmetric limit of the SIAM ($n_f=n_d=1$) for the comparison since the results for the SIAM are
universal and hence independent of the filling.

Fig.~1 shows the $f$ density of states  for the PAM and SIAM near the Fermi energy $E_F=0$. Both models show temperature-dependent Kondo peaks of similar width, but the rate of 
change of the peak intensity with temperature is much smaller  and the Kondo peak persists up to 
much higher temperatures in the PAM than for the SIAM. This is consistent
with  the protracted behavior of screened 
moments in the PAM \cite{niki2}, i.e., the rate of change of the screened moments with temperature is much smaller in the PAM compared to the SIAM when
$n_d \alt 0.8$ and $n_f \approx 1$. 
Note the difference in scales of the two parts of this figure, both the 
intensity and the spectral weight of the Kondo peak are larger in the PAM 
than in the SIAM although the hybridization parameter $V$ is larger for the 
PAM in this case. This shows (and we generally find) that the height of the Kondo peak in the PAM 
does not scale like $1/V^2$ as it does in the impurity models. 
For the PAM, starting at high temperatures the Kondo resonance peak 
is located well above the Fermi energy but as we lower the temperature it shifts 
toward $E_F$.  Hence there is a spectral weight transfer into the region around
the Fermi energy as the temperature is lowered but we can see that the spectral 
weight in this region increases slower in the PAM than in the SIAM.

Fig.~2 shows the momentum dependence of the $f$ and $d$ spectral functions for the PAM along the diagonal direction of the Brillouin zone (the main conclusions
 do not depend on the chosen direction). In the large-$D$ hypercubic lattice, the zone center (corner) corresponds to  a very large negative (positive) $\epsilon_k$.   We see that near
the zone center (lowest part of the figure) there are two apparent maxima, the lower one has mostly 
$d$ character the upper one mostly $f$ character. The latter has a narrow Kondo 
like feature but would not be seen in photoemission experiment since it is located above the Fermi energy.
The quasiparticle peak (Kondo peak) near the Fermi surface starts to develop only as ${\bf k}$ moves well away from the zone center. Note that there is a  gradual transformation of this peak from a mixed $fd$-character at $\epsilon_k=0$ to an almost entirely $f$ character at and above $\epsilon_k=1.23$.  In addition to these narrow peaks the data shows a small and broad  non-dispersive peak near $\omega=-0.8$ which has mostly $f$ character. This  is a remnant of the lower unhybridized $f$ level, we were not able to resolve the peak corresponding to the upper $f$ level for these
model parameters (we  resolve this peak if we use a larger value for
$U/V^2$). 

We clarify this situation
in Fig.~3. The symbols in this figure show the position of $f$ and $d$ spectral functions maxima versus $\epsilon_k$ and the solid line shows the quasiparticle energy, calculated from the real part of the pole of the Green's functions.
 In the narrow region  above the Fermi surface  we  see an almost dispersionless band. 
The imaginary part of these poles are very large, so they correspond to very broad peaks in the spectral functions which we were not able to resolve. 
The bands above and below this region correspond to well defined peaks in the
spectral functions.  
The general features of this
band structure for the PAM persists up to very high temperatures ($T/T_0 \agt 10$). 

The imaginary part of self-energy has a maximum near the Fermi energy (not shown); its position shifts toward $E_F$  and its value approaches zero as the temperature is lowered. This suggests a Fermi-liquid like behavior for the model. Furthermore, the real part of the effective self-energy for the $d$ electrons shows a large negative slope near the Fermi energy which denotes a large effective mass for the $d$ electrons. We can use the self-energy to calculate the optical conductivity and the dc resistivity of the model (the details of calculations will appear in a future publication).  At temperatures of the order of  $T_0^{SIAM}$ the resistivity shows a log-linear behavior adjacent to a maximum,  just like what is expected from a Kondo-type resistivity for a dilute system.   As the  temperature is lowered 
toward $T_0$ the resistivity flattens. Below  $T_0$  the resistivity decreases quickly towards zero indicating the onset of coherence and the emergence of a Fermi-liquid (metallic) behavior. 

\paragraph*{Interpretation}

Some of these results are consistent with   
a simple band-formation picture. When $V=0$ the available electronic states
consist of a  $d$ band and two (doubly degenerate) local $f$ levels separated by $U$. When $V$ is
turned on, a new resonant state forms near the Fermi surface. Furthermore,  the original $d$ band mixes with the local $f$ levels and the resonant level, giving rise to a renormalized band which has $f$ character near the
renormalized $f$ level energies and has $d$ character far from them as we see in Fig.~3. 
The Kondo states near $E_F$ have mostly $f$ character indicating that the $f$ electrons themselves are involved in screening the local moments through hybridization with the $d$ band.
This is  like the situation in a single-band Hubbard model, where the electrons within the band are responsible for screening the ``local
moments'' in that band.

We can understand the emergence of the two energy scales ($T_0^{SIAM}$ and $T_0$)
and the protracted screening of the moments in the PAM using the arguments of Nozieres\cite{Nozieres}. He argued that since the screening cloud of a local magnetic moment involves  conduction electrons  within $T^{SIAM}_0/T_F$ of the Fermi surface, only a fraction of the moments $n_{eff}\approx N_d(0) T^{SIAM}_0$         
may be screened by the conventional Kondo effect. He then proposed that the
screened and unscreened sites may be mapped onto particles and holes of an 
effective Hubbard model with local Coulomb repulsion $U$. The screening clouds  hop from site to site and  effectively 
screen all the moments in a dynamical fashion. The hopping constant of this effective model is 
strongly suppressed relative to $t^*$ by the overlap of the screened and
 unscreened states. Thus the Kondo scale of the effective model becomes
 much less than $T_0^{SIAM}$. This leads to the protracted screening behavior
which is a crossover between the two regimes of Kondo screening at the higher
scale $T_0^{SIAM}$ and coherent screening at the lower scale $T_0$. In this
argument the two scales depend  on $n_{eff}$ and hence there can be no
universal relation between them.

\paragraph*{Comparison with experiment}

For the purpose of comparison with experiment we focus our attention on the data for CeBe$_{13}$ compound because it has a large enough Kondo temperature that current high resolution apparatus should be able to probe its Kondo resonant state.
CeBe$_{13}$ is a mixed-valent heavy fermion compound with a small  number of conduction electrons per site and a rather large Kondo temperature ($T_K\approx 400 K$) so the PAM with the parameters we chose for $n_d=0.4$ is suitable for comparison with experimental observations on this material ($T_0 \approx 320 K$ if 
$t^*\approx 2 eV$). A series of photoemission experiments on single crystalline 
samples of Ce compounds including CeBe$_{13}$ have suggested a temperature 
dependence for the Kondo peak which is much weaker than what one expects from 
single-impurity model calculations\cite{jjaa}, this is consistent with Fig.~1.

Photoemission spectra obtained by choosing various incident angles on  CeBe$_{13}$ crystals have shown a dispersive structure near the Fermi surface which is persistent up to temperatures of the order of $10 T_K$\cite{arko}. As we showed in Fig.~1 the PAM predicts a Kondo peak at these higher temperatures whereas the SIAM does not. Our results also show the dispersive nature of the quasiparticles as shown in Fig.~3. The photoemission observations show that the quasiparticle peak intensity near $E_F$ is a minimum at the zone center and tends to develop as ${\bf k}$ deviates from the zone center and reaches its highest value at the zone corner. This is consistent with what we see in  Fig.~2. 

\paragraph*{Conclusions}

We find that in the metallic regime where $n_d \alt 0.8$ and $n_f\approx 1$: 
i) The PAM  predicts a Kondo peak which has a weaker temperature dependence and  persists up to much higher temperatures (in units of the Kondo scale) compared to the predictions of the SIAM.  
ii) Both the intensity and the spectral weight of the Kondo peak are larger in the PAM than in the SIAM.
iii) The Kondo peak intensity does not scale like $1/V^2$ in the PAM. 
iv) The Kondo peak is dispersive in the PAM making a heavy quasiparticle band which crosses the Fermi surface. This can be interpreted as the heavy band emerging from an effective Hubbard model for the local screening clouds introduced by Nozieres. 
v) The Kondo peak below the Fermi energy starts to develop only as ${\bf k}$ deviates from the zone center, consistent with  experiment. 
vi) There are two relevant energy scales for the PAM: the onset of screening scale $T_0^{SIAM}$, and the onset of coherence scale $T_0$ which is strongly suppressed compared to the latter.  This is also consistent with Nozieres arguments.

	These results provide a consistent interpretation of the Los Alamos
photoemission experiments involving single-crystals, but not the results
of the Bell Labs and Neuchatel groups involving poly-crystals.  Thus, a
controversy persists.  However, we note that near the insulating  regime ($n_d\approx n_f\approx 1$) both the PAM Kondo scale ($T_0$) and screening rate are enhanced compared to those of the SIAM and Nozieres' arguments do not 
apply.  In the narrow region  where the system is doped away from the insulating
state, both the screened local moment and the spectra show impurity-like
temperature dependence.  It is possible that this, or a more realistic model,
including the effects of disorder, orbital degeneracy, or crystal field
effects may provide a unifying interpretation of all of the photoemission
spectra.

	We would like to acknowledge stimulating conversations with
J.\ Allen, 
A.\ Arko,
A.\ Chattopadhyay,
D.L.\ Cox,
B.\ Goodman, 
M.\ Grioni,
D.W.\  Hess, 
M.\ Hettler, 
J.\ Joyce, 
H.R.\ Krishnamurthy, 
Th.\ Pruschke, 
P.\ van Dongen 
and
F.C.\ Zhang.

Jarrell and Tahvildar-Zadeh would like to acknowledge the support of NSF 
grants DMR-9704021 and DMR-9357199.  Freericks acknowledges the support of an 
ONR-YIP grant N000149610828. Tahvildar-Zadeh and Freericks received partial
support from the Petroleum Research Fund, administered by the American Chemical
Society (ACS-PRF 29623-GB6).
Computer support was provided by the Ohio Supercomputer Center.

\begin{figure}[t]
\epsfxsize=3.375in
\epsffile{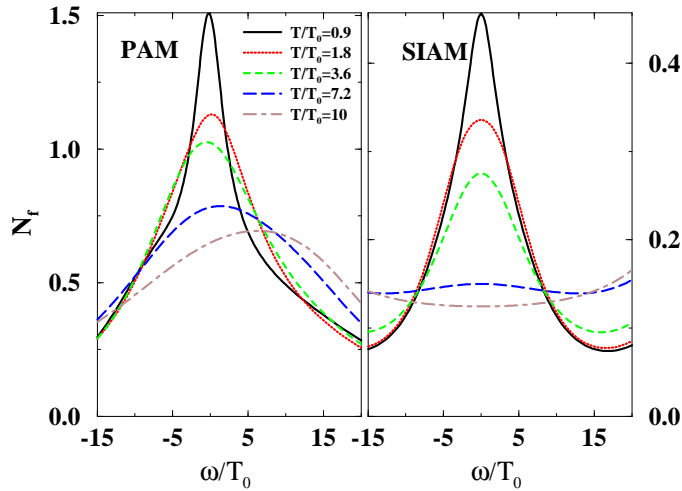}
\caption{ Near-Fermi-energy ($E_F=0$) structure of the $f$ density of states
for the asymmetric PAM (left) and the symmetric SIAM (right). The model parameters are $U=1.5$, $V=0.6$, $n_d=0.4$ for the PAM and $U=2.75$, $V=0.5$, $n_d=1.0$ for the SIAM. The temperature dependence of the peak 
is universal for the SIAM and hence independent of the band fillings. Note that $T_0$ refers to the two different Kondo scales of the PAM (left) and the SIAM (right).}
\end{figure}

\begin{figure}[t]
\epsfxsize=3.375in
\epsffile{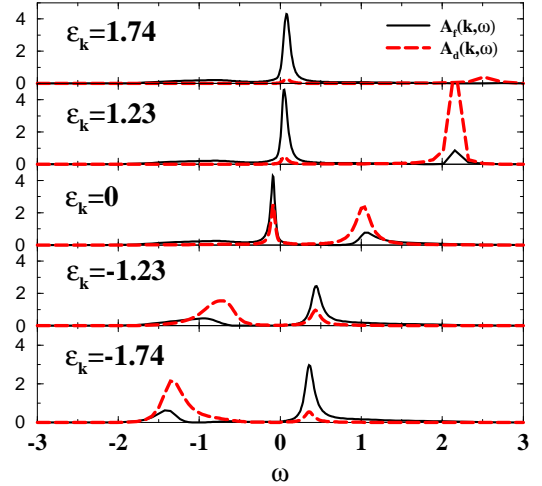}
\caption{ The PAM $f$ and $d$ spectral functions for different values of the momentum vector ${\bf k}$ for $U=1.5$, $V=0.6$, $n_f\approx 1.0$, $n_d=0.6$ and $T/T_0=0.46$. $\epsilon_k$ is the unrenormalized band energy. The Fermi energy is located at $\omega=0$.}
\end{figure}

\begin{figure}[t]
\epsfxsize=3.375in
\epsffile{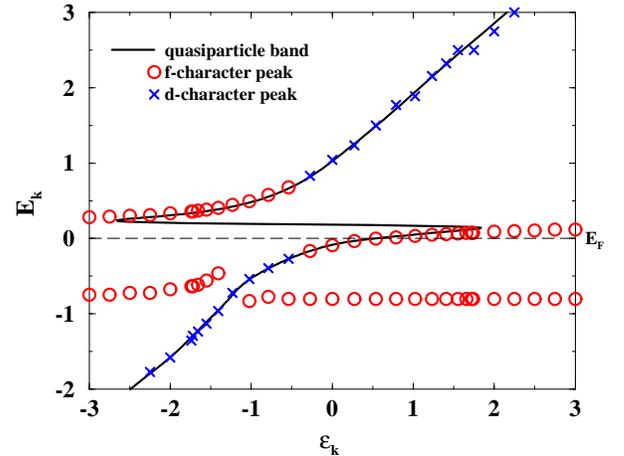}
\caption{ Band structure for the PAM. The model parameters are the same as in 
Fig.~2. However the features persist up to $T/T_0 \approx 10$. The solid line shows the real part of the Green's  functions poles vs. $\epsilon_k$ the 
unrenormalized band energy. The symbols show the positions of the maxima in the 
$f$ and $d$ spectral functions.  We characterize these peaks to be of 
$f$ character whenever $A_f(\epsilon_k,E_k)  >  A_d(\epsilon_k,E_k)$ or 
$d$ character whenever $A_f(\epsilon_k,E_k)  <  A_d(\epsilon_k,E_k)$ (cf. 
Fig.~2).}
\end{figure}

\end{document}